\documentclass[twocolumn,showpacs,amsmath,amssymb,pre]{revtex4}

\usepackage{graphicx}
\usepackage{dcolumn}
\usepackage{bm}
\usepackage{amsmath}
\usepackage{nicefrac}

\begin{document}

\title{Does ohmic heating influence the flow field in thin-layer electrodeposition?}
\author{Matthias Schr\"oter}
\email{matthias.schroeter@physik.uni-magdeburg.de}
\author{Klaus Kassner} 
\affiliation{Fakult\"at f\"ur Naturwissenschaften, Otto-von-Guericke Universit\"at Magdeburg, 
             Postfach 4120, D-39016 Magdeburg, Germany}

\author{Ingo Rehberg} 
\affiliation{Physikalisches Institut, Universit\"at Bayreuth, D-95440 Bayreuth, Germany}

\author{Josep Claret} 
\author{Francesc Sagu\'es}
\affiliation{Departament de Qu\'\i mica-F\'\i sica, Universitat de Barcelona,
             Mart\'\i  \, i Franqu\`es 1, E-08028 Barcelona, Spain}

\date{\today}

\begin{abstract}
In thin-layer electrodeposition the dissipated electrical energy leads to a substantial heating of the
ion solution. We measured the resulting temperature field by means of an infrared camera. 
The properties of the temperature field correspond closely with the development of the concentration field.
In particular we find, that the thermal gradients at the electrodes 
act like a weak additional driving force to the convection rolls driven by concentration gradients.

\end{abstract}


\maketitle

\section{Introduction}
The electrochemical deposition of metals from aqueous solutions 
in quasi-two-dimensional geometries has proven to be a valuable test bed
to examine concepts of interfacial growth like fractal growth \cite{argoul:88,kuhn:95}, 
morphological transitions \cite{fleury:91,kuhn:94,lopez:97a}
or dendritic growth \cite{barkey:95,argoul:96}. 
The properties of the evolving deposit are in many cases
sensitive to the presence of convection currents in the solution 
\cite{jorne:87,tomas:93,barkey:94,lopez:96,rosso:99,schroeter:02}. 
There are two well-known origins of convection rolls:
a) on small scales, spatial inhomogeneities of the ionic charge 
distribution can trigger electro-convection \cite{fleury:92,fleury:93,huth:95};
b) density inhomogeneities due to concentration changes at the electrodes induce 
large scale gravity-driven convection rolls 
\cite{rosso:94,barkey:94,huth:95,linehan:95,chazalviel:96,argoul:96,dengra:00,schroeter:02}. 

However there is another potential source of density changes: due to the small cell volume 
of typically $\le$ 1\,$\rm cm^3$,
the dissipated electric energy can be the source of a significant heating.
If the cell were completely thermally insulated, a solution exposed to 
an electrical power of 500\,mW would start to boil after 500\,s.
While thermal conduction will confine the overall temperature increase to smaller values, 
considerable temperature gradients might arise and generate density driven convection. 

Thermally induced convection rolls have been thoroughly studied in thin-layer geometries
heated from below (for a recent review see \cite{bodenschatz:00}) and the side 
\cite{cormack:74,imberger:74,patterson:80,boehrer:97,delgado:01}.
In contrast, the role of thermal effects was previously not examined 
in electrodeposition. 
In order to quantify the possible temperature gradients, 
a high spatial resolution of the temperature field is necessary.
We present here measurements of the temperature field at both electrodes performed by use of an
infrared camera. 
The so determined evolution of the temperature field can be related
to the development of the concentration field, which is known from interferometric
measurements \cite{argoul:96,leger:00}.
The remainder of the paper is organized as follows: 
Section~\ref{ch:setup} describes our experimental setup, 
section~\ref{ch:result} presents the measurement results, which will be discussed in 
sec.~\ref{ch:conclusions}.

\section{Experimental Setup}
\label{ch:setup}
The measurements are performed with the infrared camera Varioscan 3021-ST from InfraTec,
which contains a Stirling-cooled HgCdTe-detector with 360 $\times$ 240 pixel.
Using its macro we observe an area of 5.0 $\times$ 3.4 $\rm cm^2$, which yields a spatial resolution 
of 140\,$\mu$m. The maximal image capturing frequency is 1.1\,Hz, 
the thermal resolution $\pm$ 30\,mK. 

\begin{figure}[htbp]
  \begin{center}
    \includegraphics[width=8.6cm]{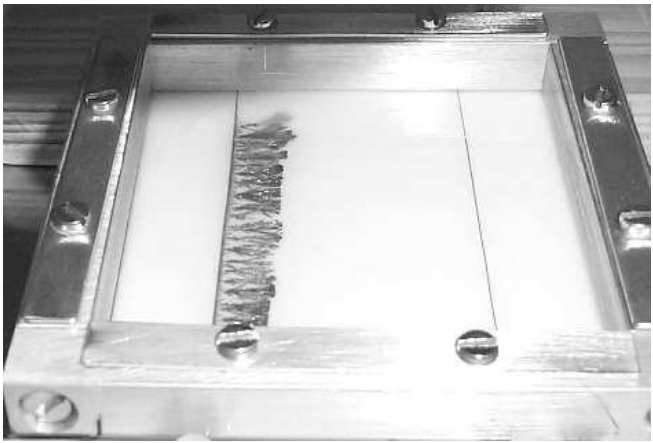}
    \caption{Cell used for the infrared measurements. The bottom (white) consists of Teflon, 
        the upper cover is a polyethylene foil stretched over an aluminium frame. 
        The electrodes are parallel zinc wires with a distance of 4\,cm.
        The deposit at the cathode has grown for 410\,s.}
    \label{fig:setup}
  \end{center}
\end{figure}

While the sensor is sensitive for wavelengths in the range of 8-12 $\mu$m, 
glass plates, which are normally used as top and bottom plate of the cell, are opaque
in this region.  Therefore  
the upper cell cover was realized using a polyethylene foil stretched over an aluminium
frame. 
Due to the flexibility of the polyethylene foil and a small overpressure necessary to fill the cell,
the plate separation of about 0.5 mm is not well defined.
The bottom plate of the cell consists of a block of Teflon, because this material has 
a low reflectivity in the infrared. This reduces the so called narcissism:
the response of the cooled detector to its own mirror image.  
The electrodes are parallel zinc wires (Goodfellow 99.99\%)  of 0.25\,mm diameter and separated by a distance of 4\,cm.
Fig.\ref{fig:setup} shows an image of the cell during an experiment.

The cell is filled with an 0.1\,M $\rm ZnSO_4$ solution  prepared from Merck p.a.\ chemicals 
in nondeaerated ultrapure $\rm H_2O$. The measurements are performed at a constant potential
of 20 $\pm$ 0.003 V. Due to the current increase during the electrodeposition process, 
the average electrical power feed $\dot{Q}_{\rm el}$ increases from 470\,mW 
at the beginning of the experiment to 650\,mW after 500\,s.

Because of the modified cell construction, the question of comparability  
with experiments performed in standard electrodeposition cells 
could be raised. Therefore we calculate the heat-flux $k_i A_i$ for the confining
plates, where  $A_i$ and $k_i$ are the area and the heat transfer coefficient of the plate.
It is important to keep in mind, that $1/k_i$
is equivalent to $1/\alpha_1 + 1 / \alpha_2 + \Delta z_i / \lambda_i$. 
Here $\alpha_1$ and $\alpha_2$ are the heat transition numbers from solution to plate resp.\ plate to air,
$\lambda_i$ is the heat conductivity and $\Delta z_i$ the thickness of the plate. 

Inserting the material
parameters of our setup~\cite{james:92,lindner:89} and using only the area between the two electrodes,
we derive a $k_i A_i$ of 16.2\,mW per Kelvin temperature difference
for the polyethylene foil and 11.5\,mW/K for the Teflon plate,
while a typical glass plate (Schott BK 7, 6.3\,mm thick) yields 16\,mW/K. 
So in a first approximation our setup is thermally equivalent to a standard electrodeposition cell.

\section{Experimental Results}
\label{ch:result}
Fig.~\ref{fig:setup} shows the growing deposit at the cathode, which belongs to the 
homogenous morphology~\cite{sagues:00}.
It is characterized by tip-splitting but retaining a growing front parallel to the cathode. 
After 500\,s a Hecker transition~\cite{kuhn:94} takes place
and the growth front breaks up into more spatially localized zones of active development.

\begin{figure}[htbp]
  \begin{center}
    \includegraphics[width=8.6cm]{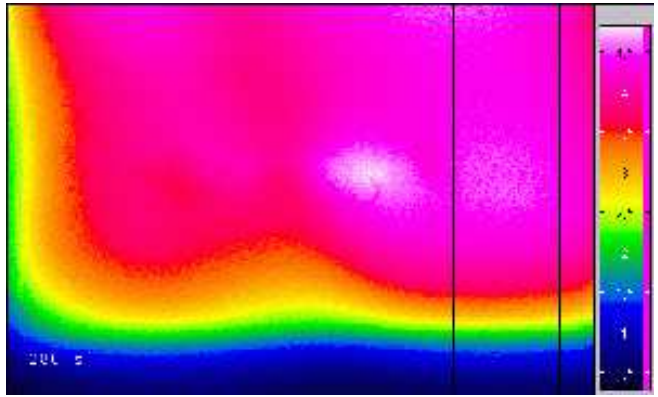}
    \caption{Thermography taken 280\,s after the start of the experiment. 
        The size of the image is 5.0 $\times$ 3.4 $\rm cm^2$.
        The white line corresponds 
        to the position of the anodic zinc wire, the rows between the black lines are averaged to
        produce Fig.~\ref{fig:T_dev_a}.}
    \label{fig:thermo}
  \end{center}
\end{figure}

Immediately before each experiment the infrared camera takes a zero image, which is then subtracted
from all images taken during the experiment. Therefore the thermographies solely depict
the temperature increase.
Fig.~\ref{fig:thermo} gives an example of such a thermography after 280\,s, the scale at the right
describes the temperature increase with respect to the beginning of the experiment.
The white line marks the position of the anodic zinc wire. 
The temperature decrease at the left hand side of the image is caused by the thermal conductivity 
of the aluminium frame, the warm ``island'' in the middle is due to the inhomogeneity of the cell thickness.

In order to improve the signal to noise ratio, 
zones of spatial homogeneity and a width of 9.1\,mm are chosen by visual
inspection. Such a zone is depicted in Fig.~\ref{fig:thermo} with two parallel black lines.
Inside this zone all rows are averaged, yielding a temperature increase $\Delta T (y,t)$,
which is only a function of distance to the cathode $y$ and time.

\begin{figure}[htbp]
  \begin{center}
    \includegraphics[angle=-90,width=8.6cm]{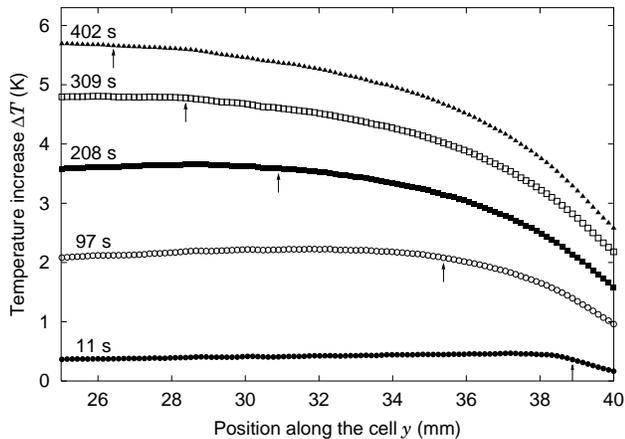}
    \caption{Temporal evolution of the temperature field at the anode,
        which is located at 40\,mm. 
        All data are averaged over a width of 9.1\,mm.
        The arrows indicate 
        the position where the deviation of $\Delta T$ from the bulk temperature
        becomes less than 20\,mK.}
    \label{fig:T_dev_a}
  \end{center}
\end{figure}

Fig.~\ref{fig:T_dev_a} shows the evolution of $\Delta T$ in the neighborhood of the anode.
It is clearly visible that the temperature increase at 
the anode itself lags behind with respect to the one observed in the middle of the cell.

\begin{figure}[htbp]
  \begin{center}
    \includegraphics[angle=-90,width=8.6cm]{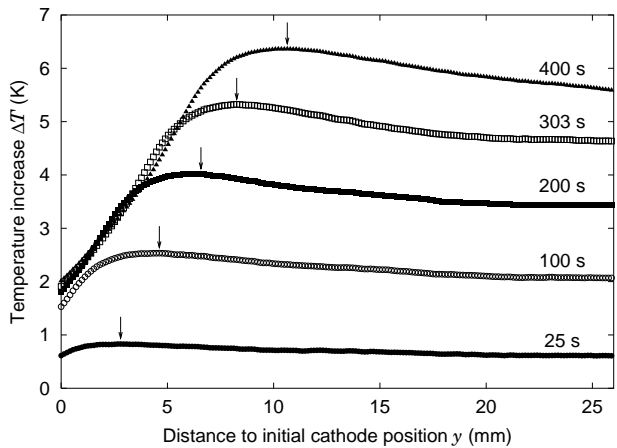}
    \caption{Temporal evolution of the temperature field at the cathode. 
        All data are averaged over a width of 9.1\,mm.
        The arrows mark the position of the temperature maximum.}
    \label{fig:T_dev_k}
  \end{center}
\end{figure}

Fig.~\ref{fig:T_dev_k} illustrates the evolution of the temperature field at the cathode.
The most prominent feature is the existence of a local temperature maximum, denoted with small
arrows. This maximum moves  towards the middle of the cell, where a plateau of spatially constant 
temperature is located.

\begin{figure}[htbp]
  \begin{center}
    \includegraphics[angle=-90,width=8.6cm]{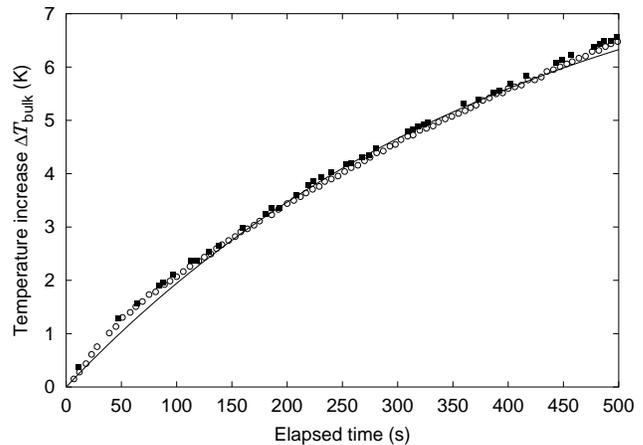}
    \caption{Temperature increase in the bulk of the cell at a distance of 14.4\,mm to the anode and
        25.6 mm\,to the cathode.
        The $\circ$ correspond to the experiment presented in Fig.~\ref{fig:T_dev_k}
        the {\tiny $\blacksquare$} to the one in Fig.~\ref{fig:T_dev_a}.
        The solid line is a fit of Equation~\ref{eq:T_dev} to the $\circ$.}
    \label{fig:bulk}
  \end{center}
\end{figure}

In order to characterize the heating process in the cell, we pick the temperature at
a distance of 25.6\,mm to the cathode (this is approximately the location, 
where the two convection rolls emerging from the electrodes finally  meet, 
as will be discussed in Section~\ref{ch:conclusions_c}).  
Referring to this point, we will speak of the {\it bulk}\/ in the following.  
In Fig.~\ref{fig:bulk} this temperature increase $\Delta T_{\rm bulk}$ is given for the 
two experiments presented in Fig.~\ref{fig:T_dev_a} and~\ref{fig:T_dev_k}.
The fact that the two measurements are almost identical  reflects the reproducibility
of the experiment. The solid line is a fit to an exponential function, which is motivated in detail
in Section~\ref{ch:conclusions_b}.  

\begin{figure}[htbp]
  \begin{center}
    \includegraphics[angle=-90,width=9.5cm]{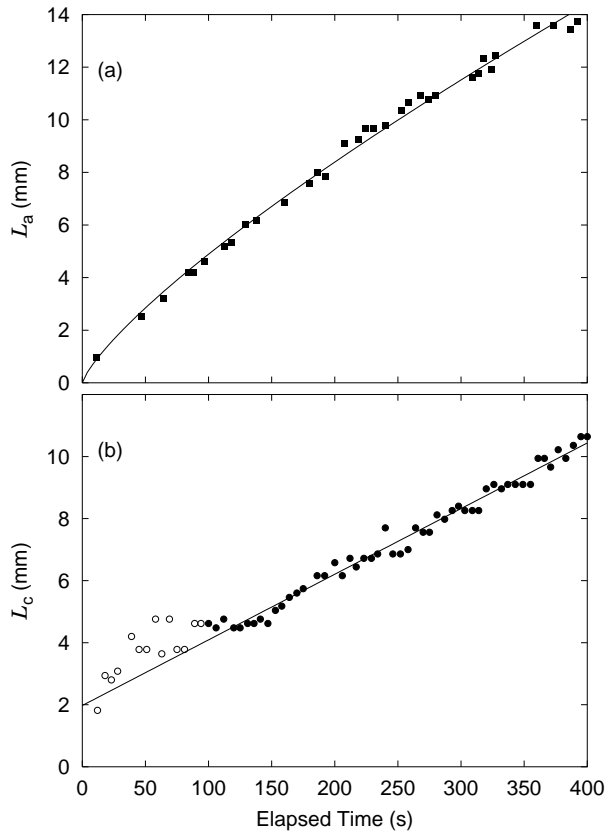}
    \caption{Temporal evolution of: (a) the distance between the anode and the point,
        where the temperature starts to deviate from the bulk.
        The solid line is a fit with Equation~\ref{eq:L_a}.
        (b) the distance between the temperature maximum and 
        the initial cathode position.
        The fit is performed with Equation~\ref{eq:L_c} to the data indicated
        with filled symbols.}
    \label{fig:L_t}
  \end{center}
\end{figure}

The arrows in Fig.~\ref{fig:T_dev_a} represent the location, where the temperature
is 20 mK\,smaller than $\Delta T_{\rm bulk}$. The distance of these points with respect to the
anode is denoted $L_a(t)$ and is shown in Fig.~\ref{fig:L_t} (a). 
Its monotonous increase with time is fitted by a power law, which will be  explained 
in Section~\ref{ch:conclusions_a}.  
Correspondingly, Fig.~\ref{fig:L_t} (b) shows the growth of the 
distance $L_c(t)$ between the the location of the temperature maximum in Fig.~\ref{fig:T_dev_k} and the cathode.
The straight line is a fit to all data with $t >$ 100\,s, which will be motivated in Sec.~\ref{ch:conclusions_c}.

\section{Discussion and conclusions}
\label{ch:conclusions}
In principle, the energy balance involves three contributions: the electrical energy feed
into the cell, the dissipated ohmic heat and the chemical reaction 
energy. However the short calculation presented in the 
appendix supports the assumption that the chemical energy contribution is mostly irrelevant.

The ohmic heat dissipated at some position in the cell will be proportional to the local 
resistivity $\rho(y,t)$ in a one-dimensional model,  
while $\rho$ will depend on the local ion concentration $c(y,t)$.
In the next three subsections we will compare the evolution of the temperature field
at the anode, at the cathode and in the bulk with the development of the 
concentration field, which is known from interferometric
measurements \cite{argoul:96,leger:00}.

\subsection{Evolution of the temperature field at the anode}
\label{ch:conclusions_a}
At the anode $\rm Zn^{2+}$ ions go  into solution, increasing the local concentration and 
therefore density. While this denser
solution sinks down to the bottom plate, it gets replaced by less dense bulk solution.
This mechanism drives a convection roll of size $L$
\cite{huth:95,chazalviel:96,schroeter:02}.
According to the fluid dynamical description there are two growth regimes:
Initially during the so called immiscible fluid regime $L$ will grow with $t^{0.8}$, 
while after some time there will be a crossover to the diffusion hindered spreading regime
with a  $t^{0.5}$ growth law.
Chazalviel et al.~\cite{chazalviel:96} showed, that both the fluid velocity $v$ and 
the ion concentration  decrease with increasing distance to the electrode.
The distance, where $v$ has dropped to zero and coincidently $c$ has fallen to $c_{\rm bulk}$,
defines $L$. 

Figure~\ref{fig:T_dev_a} shows that the temperature increase in the region between the anode and the arrows 
lags behind the one observed in the bulk.
As $c > c_{\rm bulk}$ translates into reduced ohmic heating, we identify the position of the arrows
with the end of the anodic convection roll. A fit to its length with:
\begin{equation}
    \label{eq:L_a}
        L_{\rm a}(t) =  a t^{b}
\end{equation} 
is presented in Fig~\ref{fig:L_t} (a). It yields:
a  = 0.13 $\pm$ 0.01 mm  and b = 0.78 $\pm$ 0.01, which indicates that the convection
roll is in the immiscible fluid regime for the whole run of the experiment.
According to the scaling analysis presented in Ref.~\cite{schroeter:02},
this corresponds to an average plate separation of about 650\,$\mu$m.

\subsection{Evolution of the temperature field at the cathode}
\label{ch:conclusions_c}
Due to the growing deposit, there is a zone of ion depletion in the vicinity of the cathode,
the size of which will be affected by the convection roll driven by the occuring 
density difference. As the decreased $c$ leads to higher dissipation,
the increased temperature denoted by the arrows in Fig.~\ref{fig:T_dev_k} 
is qualitatively explained. 
The distance of the arrows to the initial cathode position $L_{\rm c} (t)$ 
should correspond to the actual size of the deposit. Therefore 
we perform a linear fit with:
\begin{equation}
    \label{eq:L_c}
        L_{\rm c}(t) = v_{\rm c} t + l_{\rm c}
\end{equation}
which is shown in Fig.~\ref{fig:L_t} (b). We derive 
$v_{\rm c}$ = 21.2  $\pm$ 0.4  $\mu$m/s, which agrees well with  20.7 $\pm$ 0.8 $\mu$m/s   
front velocity  determined from photographs of the deposit.

$l_{\rm c}$ is found to be 2 $\pm$ 0.1 mm.  
This finite distance can be explained by the absence of heat production in the 
metallic deposit because of its
low resistivity and the fact that its heat conductivity  is 190 times higher than water. 
So the deposit is an effective heat sink, the
resulting heat flux shifts the temperature maximum into the cell.

If the cathodic convection roll, apart from the fact that it starts at the actual
front of the deposit, grows in the same way as the anodic roll, 
they  meet  450\,s after the beginning of the experiment at a distance of 25\,mm to the cathode.  
This corresponds to the observed change in morphology after that time.

\subsection{Temperature evolution in the bulk}
\label{ch:conclusions_b}
The temperature increase observed in Fig.~\ref{fig:bulk}
can be modeled, if we assume, that the whole cell shares the
constant bulk properties $c$ and $\rho$ and therefore $T$. 
The  supplied electrical power $\dot{Q}_{\rm el}$ would then be compensated by the  
heating of the system with heat capacity $C$ and the
heat flow  $\dot{Q}_{\rm flow}$:
\begin{equation}
   \dot{Q}_{\rm flow} =  -  (T - T_0) \sum_i k_i A_i  
  \label{eq:waermeleit}
\end{equation}
where $T_0$ represents the ambient temperature and $T$ the temperature inside the electrolyte.
If we assume $\dot{Q}_{\rm el}$ to be constant, the corresponding differential equation:
\begin{equation}
   \dot{Q}_{\rm el} =  (T - T_0) \sum_i k_i A_i + C \frac{\partial T} {\partial t}  
  \label{eq:waermeleit_dgl}
\end{equation}
has the straightforward solution:
\begin{equation}
  \label{eq:T_dev}
  (T - T_0) = T_{\rm final} \left(1 - e^{\nicefrac{- t}{\tau}}  \right) 
\end{equation}
Here $T_{\rm final} = \dot{Q}_{\rm el}/ \sum_i k_i A_i$ denotes the finally reached temperature difference
and $\tau = C / \sum_i k_i A_i$ is the time constant of the heating up. 
A fit of Equation~\ref{eq:T_dev} to the experimental data is displayed in Fig.~\ref{fig:bulk}. 
It yields:
$T_{\rm final}$ = 9 $\pm$ 0.2 K. 
This translates to an overall heat-flux $\sum_i k_i A_i$ of 60\,mW  per K temperature difference.
A comparison of this result with the values of $k_i A_i$ calculated in Sec.~\ref{ch:setup} shows that 
about 50\% of the heat flux takes place through the top and the bottom plate of the cell. 
The heat flux through the side walls, the electrodes 
and the plate area beyond the electrodes accounts for the rest.

\subsection{Influence of the temperature gradients on the  convection rolls} 

\begin{figure}[htbp]
  \begin{center}
    \includegraphics[angle=-90,width=8.6cm]{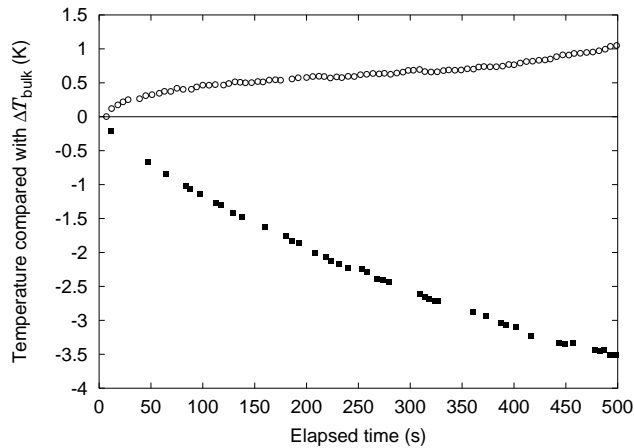}
    \caption{Temperature difference between the bulk and $\circ$: the cathodic 
        temperature maximum  and {\tiny $\blacksquare$}: the temperature at the anode.}
    \label{fig:Delta_T}
  \end{center}
\end{figure}

In order to estimate the influence of the temperature gradients on the concentration
driven convection currents,
we plot in Fig.~\ref{fig:Delta_T} the difference between  $\Delta T_{\rm bulk}$ and 
a) the cathodic temperature maximum and b) the anode. 
These temperature gradients result in a down-flow at the anode and an up-flow
at the growth front, so they act as an additional driving.

\begin{figure}[htbp]
  \begin{center}
    \includegraphics[angle=-90,width=8.6cm]{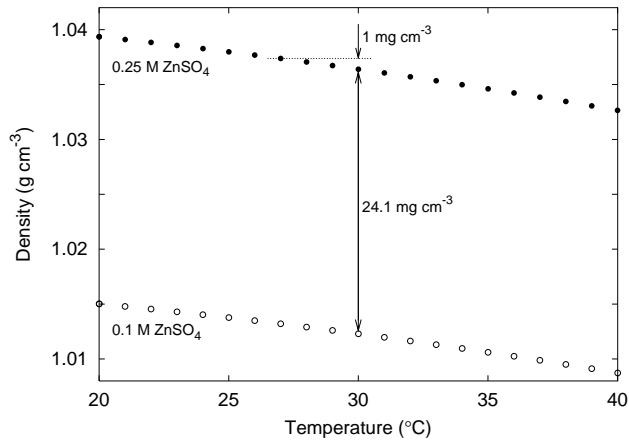}
    \caption{Temperature dependency of the densities at the anode and in the bulk. The
        $\bullet$ correspond to $c$ = 1.0\,m $\rm ZnSO_4$, the $\circ$ to 
        $c_{\rm bulk}$ = 0.1\,m $\rm ZnSO_4$.}
    \label{fig:rho_diff}
  \end{center}
\end{figure}

For a more quantitative determination of their contribution, the temperature dependency of the 
density was measured for different concentrations. The applied
density measurement instrument DMA 5000 from Anton Paar has an accuracy better than
50\,$\mu$g/$\rm cm^3$. 

After 400\,s the temperature of the 0.1\,M $\rm ZnSO_4$ bulk solution has reached 
about 30\,$^\circ$\,C which corresponds to a density of 1012.3 mg/$\rm cm^3$.
In accordance with the measurements presented in Ref.~\cite{argoul:96} and the theory 
in Ref.~\cite{chazalviel:96}, we estimate $c$ at the anode for that time as 0.25\,M. 
This corresponds to a density of  1036.4\,mg/$\rm cm^3$ for $T$ = 30\,$^\circ$\,C and
1037.4\,mg/$\rm cm^3$ for the actually measured $T \approx$ 27\,$^\circ$\,C.
So the contribution of the temperature gradient to the overall density difference at the
anode is about 4\% as visualized in Fig.~\ref{fig:rho_diff}.

At the cathode $c$ has reached zero at that time, $\rho$ of $\rm H_2O$ is 995.68\,mg/$\rm cm^3$
for $T$ = $30^\circ$\,C and 995.38\,mg/$\rm cm^3$ for the $T$ = $31^\circ$\,C at the maximum.
This results in a 2\% contribution of the temperature gradient to the total density contrast.

As our applied potential is above average for standard electrodeposition experiments, we conclude that
temperature inhomogeneities will only weakly contribute to the density driven convection rolls.
This result justifies with hindsight the use of the theoretical description of Chazalviel et 
al.\cite{chazalviel:96}.
Finally it should be remarked that morphologies like stringy \cite{trigueros:92}, 
where the zone of active growth is restricted to few small spots with very high
local current densities, may differ substantially from our results.

\begin{acknowledgments}
We want to thank J\"urgen Fiebig from InfraTec for his friendly support.
We are also grateful to Niels Hoppe and Gerrit Sch\"onfelder from IMOS, Universit\"at Magdeburg
for their assistance with the density measurements.
This work was supported by the {\it Deutsche Forschungsgemeinschaft}\/ under the project
FOR 301/2-1. Cooperation was facilitated by the TMR Research Network FMRX-CT96-0085:  
Patterns, Noise \& Chaos.
\end{acknowledgments}

\section{appendix}

The standard enthalpy of formation $ \Delta_{\rm b} H^{\ominus}$
of $\rm Zn^{2+}$ ions in an infinitely diluted solutions is $-153.89$\,kJ/mol \cite{atkins:90}. 
It contains three different contributions:
1) the energy necessary to liberate an atom from the surrounding lattice, 
2) the energy needed to ionize the atom and 
3) the hydration energy, which is set free when the water dipoles surround the ion.
Only the last term depends on the concentration of the solution, 
decreasing with the number of ions already dissolved.

As the reaction rate at the electrodes is directly proportional to the electrical current $I$, 
so is the chemical power $\dot{Q}_{\rm chem}$ :
\begin{equation}
  \label{eq:Q_chem}
   \dot{Q}_{\rm chem} = \frac{\Delta_{\rm b} H^{\rm \ominus}}{F \;z} \; I
\end{equation}
where $F$ is the Faraday constant (96485\,As/mol) and $z$ the charge number.
For zinc we derive $\dot{Q}_{\rm chem}/ I$ =  -0.8\,mW/mA. In the vicinity of the anode
this number leads to additional heating due to the transformation of chemical energy.
At the cathode the chemical energy stored in the newly produced zinc has to be
subtracted from the overall heat production, but does not lead to a 
direct cooling of the cathode.

With our experimental conditions 
an average current of 30\,mA feeds an electrical power of 600\,mW  into the cell.
Compared to that  $\dot{Q}_{\rm chem}$ is 24\,mW which is about 4\% of the electrical power.

At the cathode the ion concentration drops  fast to zero, 
which justifies the approximation of an infinitely diluted solution.
At the anode the ion concentration increases during the whole run of the experiment. 
Therefore the hydration energy of the newly produced  $\rm Zn^{2+}$ ions decreases and 
in consequence the heat, which is released from the chemical reaction decrease as well.
So the overall heat production from chemical energy at the anode is smaller than 
estimated here and therefore mostly irrelevant for the energy balance.
This conclusion agrees with the fact that the anode is the coldest point of the cell 
as clearly shown in Fig~\ref{fig:T_dev_a}, in spite of this extra heat production there.

\end{document}